\documentclass[a4paper]{cas-dc}
\usepackage[numbers]{natbib}
\usepackage{makecell, multirow}
\usepackage{float}
\usepackage{placeins}
\usepackage{url}

\def\tsc#1{\csdef{#1}{\textsc{\lowercase{#1}}\xspace}}
\tsc{WGM}
\tsc{QE}

\hypersetup{draft}
\begin{document}
\let\WriteBookmarks\relax
\def\floatpagepagefraction{1}
\def\textpagefraction{.001}
\let\printorcid\relax 

\shorttitle{}    

\shortauthors{Yuchen Qi et al.}

\title [mode = title]{VB-NET: A physics-constrained gray-box deep learning framework for modeling air conditioning systems as virtual batteries}  

\author[1]{Yuchen Qi}
\author[2]{Ye Guo}
\cormark[1]
\author[1]{Yinliang Xu}

\address[1]{Shenzhen International Graduate School (SIGS), Tsinghua University, Shenzhen 518071, China} 
\address[2]{Department of Building Environment and Energy Engineering, The Hong Kong Polytechnic University, Kowloon, Hong Kong, China}
\cortext[1]{The work was supported in part by the National Natural Science Foundation of China under Grant 52377105. (Corresponding author: Ye Guo, e-mail: ye.guo@polyu.edu.hk)}  

\begin{abstract}
The increasing penetration of renewable energy necessitates unlocking demand-side flexibility. While air conditioning (AC) systems offer significant thermal inertia, existing physical and data-driven models struggle with parameter acquisition, interpretability, and data scarcity. This paper proposes VB-NET, a physics-constrained gray-box deep learning framework that transforms complex AC thermodynamics into a standardized Virtual Battery (VB) model. We first mathematically prove the isomorphic equivalence between the AC and VB models. Subsequently, VB-NET is designed to strictly enforces physical laws by decoupling shared meteorological drivers from private building thermal fingerprints and embedding a differentiable physics layer. Experimental results demonstrate that VB-NET significantly outperforms conventional black-box models in state of charge tracking while successfully recovering underlying thermodynamic laws to yield physically consistent parameters. Furthermore, utilizing multi-task learning and terminal sensitivity modulation, VB-NET overcomes the cold-start dilemma, achieving high-precision modeling for new AC units using only 2\% to 6\% of historical data. Ultimately, this study provides an interpretable and data-efficient pathway for aggregating decentralized AC resources for grid regulation.
\end{abstract}



\maketitle

\section{Introduction}
\label{intro}
The global energy landscape is undergoing a profound transformation aimed at achieving carbon neutrality \cite{sepulveda2021design}. As the penetration of renewable energy sources deepens, the inherent uncertainty on the supply side exacerbates the flexibility mismatch between power generation and consumption \cite{negrete2017value}. Consequently, excavating flexibility resources on the demand side is imperative to maintain a continuous power balance. Globally, buildings---specifically air conditioning (AC) systems---account for approximately 20\% of total electricity consumption \cite{perez2008review}, due to their significant thermal inertia, AC systems possess the potential to function as valuable flexible resources. To fully exploit this potential, precise modeling of AC characteristics is a prerequisite.

Thermodynamic Equivalent Thermal Parameter (ETP) models represent the most prevalent approach for characterizing AC systems \cite{sonderegger1978dynamic, taylor1988effects}. Grounded in electrical circuit analogies, these models map the thermal dynamics of building spaces to electrical components, where thermal resistance ($R$) reflects the heat transfer properties of the envelope (e.g., walls and windows) \cite{fux2014ekf}, and thermal capacitance ($C$) represents the heat storage capacity of indoor air and mass \cite{berthou2014development}. Selecting the order of an ETP model involves a trade-off between fidelity and computational complexity; while high-order models quantify heat transfer with greater precision, they incur significant computational costs \cite{liu2015model, wang2020flexibility}. Therefore, first- or second-order simplified structures are typically adopted to balance accuracy and efficiency \cite{vrettos2016experimental, pan2017feasible}.

Although thermodynamic ETP models explicitly describe the relationships between operating power and indoor/outdoor temperatures, they face a critical practical hurdle: thermodynamic parameters ($R$ and $C$) are difficult to measure directly and exhibit high heterogeneity across different buildings \cite{olu2022building}. This dependence on hard-to-obtain physical parameters severely restricts the scalability of ETP models in large-scale demand-side regulation \cite{mahdavi2016mapping}.

To circumvent the reliance on physical parameters, researchers have increasingly turned to data-driven methodologies. Traditional machine learning algorithms, such as Support Vector Machines (SVM) \cite{wei2018review}, Random Forests (RF) \cite{amasyali2018review}, and Artificial Neural Networks (ANN) \cite{ahmad2017trees}, have been widely employed to abstract the non-linear dynamics of AC systems. With the proliferation of operational data, deep learning techniques, including Recurrent Neural Networks (RNN) \cite{lv2022building}, Temporal Convolutional Networks (TCN) \cite{he2024short} and Long Short-Term Memory networks (LSTM) \cite{wang2023physics}, have been adopted to construct black-box models from historical records. To further enhance performance, hybrid frameworks incorporating such as expert systems \cite{gao2025deep}, transfer learning \cite{liu2022hybrid}, and genetic algorithms \cite{zhang2021review} have also been explored.

Despite the flexibility offered by data-driven approaches, significant challenges remain. First, most existing methods function as black boxes, lacking the interpretability required for reliable engineering applications. Second, current research predominantly focuses on load forecasting rather than modeling the energy storage capabilities (i.e., thermal inertia) of AC systems, failing to fully characterize their flexibility. Finally, given the vast number of dispersed, small-scale AC units, existing methods struggle to establish precise, individualized models for each unit efficiently.

To address these gaps, there is an urgent need for a modeling framework that seamlessly integrates AC systems with the power grid. Leveraging the energy storage characteristics inherent in thermal inertia, this study proposes a novel data-driven framework that transforms the AC model into a Virtual Battery (VB) model, enabling direct participation in demand-side regulation. The specific contributions of this paper are as follows:

\begin{itemize}
    \item We propose a methodology to transform AC building models into virtual battery models. This approach fully characterizes the energy storage and dynamic properties of AC systems, demonstrating the feasibility of seamless grid integration and direct load regulation.
    
    \item We utilize deep learning to identify the specific parameters of the virtual battery, eliminating the dependency on physical parameters that are difficult to measure in practice.
    
    \item We design a specialized deep learning model, VB-Net, tailored for this task. Its key advantages include: (i) decoupling and encoding shared environmental features and private unit characteristics to achieve cold start capability for new AC units with limited data; (ii) separating dynamic and static parameters within the virtual battery to ensure physical consistency; and (iii) incorporating a physics-differentiable evolution layer to guarantee that the slope and magnitude of predicted curves strictly adhere to the dynamic laws of the virtual battery.
\end{itemize}

The remainder of this paper is organized as follows: Section \ref{sec:problem_modeling} formulates the physical models and defines the state mapping from the AC system to the virtual battery. Section \ref{sec:theoretical_analysis} provides the theoretical analysis and rigorous mathematical proof demonstrating the isomorphic equivalence of the two models. Section \ref{network} details the architecture of the proposed VB-Net. Section \ref{case_study} presents the case study and validates the effectiveness of the proposed method.Section \ref{sec:conclusion} concludes the paper.

\section{Problem Modeling}
\label{sec:problem_modeling}

To harness the demand-side flexibility of AC systems, it is essential to translate their thermal dynamics into a standardized energy storage format that can be seamlessly recognized and dispatched by the power grid. In this section, we formulate the physical models of the AC system and the proposed VB, and define the mapping mechanism between the thermal and electrical domains. Without loss of generality, the following formulation focuses on the cooling mode of the AC system during summer operations.

Consider a thermal zone regulated by an AC unit. The temporal evolution of the indoor temperature $T(t)$ is governed by the first law of thermodynamics, which dictates the energy balance within the building envelope:
\begin{equation}
    C_{th} \frac{dT(t)}{dt} = \mathcal{Q}(t, T(t), \mathbf{x}_{env}(t)) - \eta P_{ac}(t),
    \label{eq:thermo_dyn}
\end{equation}
where $T(t) \in [T_{min}, T_{max}]$ denotes the indoor temperature constrained by user comfort settings. $C_{th} \in \mathbb{R}^+$ represents the equivalent thermal capacitance of the building mass, which is assumed to be constant within the specific operating range. $P_{ac}(t)$ is the electrical power input of the AC unit, and $\eta$ denotes the coefficient of performance (COP). The term $\mathcal{Q}(\cdot)$ represents the generalized net heat gain function, quantifying heat transfer from the environment (e.g., ambient temperature, solar radiation) and internal thermal loads. It is crucial to note that we do not assume $\mathcal{Q}$ to be linear; it comprehensively encompasses the complex, non-linear thermal dynamics of the building envelope.

Parallel to the thermodynamic model, we propose modeling the AC system's flexibility as a Virtual Battery. This approach abstracts the complex building thermodynamics into a set of standard electrical energy storage parameters. We define a VB model whose State of Charge (SOC), denoted as $S(t)$, evolves according to the following dynamic equation:
\begin{equation}
    \frac{dS(t)}{dt} = \frac{1}{C_f} \left[ \eta P_{ac}(t) - P_{loss}(t) \right],
    \label{eq:vb_dyn}
\end{equation}
where $S(t) \in [0, 1]$ represents the normalized energy level of the virtual battery. $C_f$ is the virtual capacity (analogous to the maximum energy storage limit), and $P_{loss}(t)$ is the time-varying power loss rate representing the passive energy dissipation of the battery to the environment.

\begin{figure}[!htbp]
\centering
\includegraphics[width=0.7\linewidth]{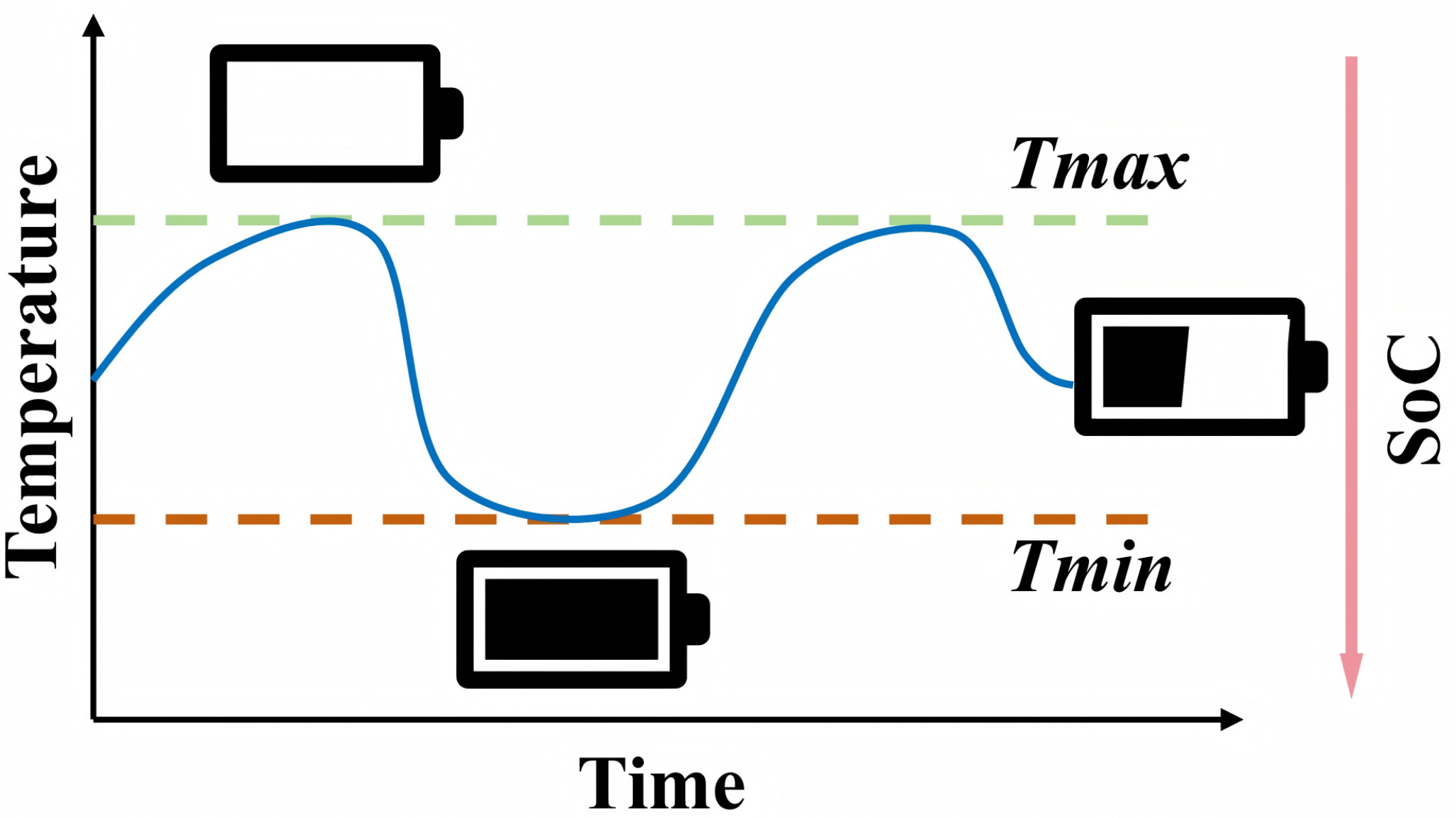}
\caption{The mapping from the indoor temperature of AC system to the SOC of VB in summer cooling case.}\label{fig:mapping}
\end{figure}

To mathematically link the thermal domain to the electrical domain, we postulate a linear inverse mapping $\Phi: [T_{min}, T_{max}] \to [0, 1]$:
\begin{equation}
    S(t) = \Phi(T(t)) = \frac{T_{max} - T(t)}{T_{max} - T_{min}}.
    \label{eq:mapping}
\end{equation}
This mapping normalizes the heterogeneous temperature preferences of different users into a unified metric. As illustrated in Fig.~ \ref{fig:mapping}, this mapping implies that $S(t)=1$ (fully charged state) corresponds to $T_{min}$ (maximum stored cooling energy), while $S(t)=0$ (depleted state) corresponds to $T_{max}$ (minimum stored cooling energy). 

\section{Theoretical Analysis}
\label{sec:theoretical_analysis}

While the state mapping proposed in Section \ref{sec:problem_modeling} provides an intuitive conceptual bridge, it is imperative to establish a rigorous mathematical foundation to guarantee physical fidelity. In this section, we provide a mathematical derivation demonstrating that the thermodynamic AC model and the VB model are strictly equivalent under specific parameter conditions. Based on the formulations above, we establish the following theorem regarding the equivalence of the two systems.

\noindent \textbf{Theorem.} Considering the thermodynamic AC system described by Eq. \eqref{eq:thermo_dyn} and the mapping $\Phi$ defined in Eq. \eqref{eq:mapping}, there exists a pair of parameters $\{C_f, P_{loss}(t)\}$ such that the trajectory of the virtual battery state $S(t)$ is mathematically identical to the mapped temperature trajectory $\Phi(T(t))$ for any control input $P_{ac}(t)$. Furthermore, $C_f$ is a time-invariant constant, whereas $P_{loss}(t)$ is a time-variant parameter.

\noindent \textit{Proof.} First, we differentiate the mapping function Eq. \eqref{eq:mapping} with respect to time $t$. Let the temperature deadband be denoted as $\Delta T = T_{max} - T_{min}$. Since the comfort bounds $T_{max}$ and $T_{min}$ are constant user settings, we obtain:
\begin{equation}
    \frac{dS(t)}{dt} = -\frac{1}{\Delta T} \frac{dT(t)}{dt}.
    \label{eq:diff_map}
\end{equation}
Substituting the thermodynamic governing equation Eq. \eqref{eq:thermo_dyn} into Eq. \eqref{eq:diff_map} yields:
\begin{equation}
    \frac{dS(t)}{dt} = -\frac{1}{\Delta T} \left[ \frac{1}{C_{th}} \left( \mathcal{Q}(t, T(t), \mathbf{x}_{env}(t)) - \eta P_{ac}(t) \right) \right].
\end{equation}
Rearranging the terms to align with the structure of the battery dynamics gives the derived dynamic equation:
\begin{equation}
    \frac{dS(t)}{dt} = \frac{1}{C_{th} \Delta T} \left( \eta P_{ac}(t) - \mathcal{Q}(t, T(t), \mathbf{x}_{env}(t)) \right).
    \label{eq:derived_dyn}
\end{equation}
For the virtual battery model Eq. \eqref{eq:vb_dyn} to perfectly describe the dynamics of Eq. \eqref{eq:derived_dyn}, the coefficients of the control input $P_{ac}(t)$ and the bias terms must be identical. Comparing the two equations yields the following identification conditions:
\begin{equation}
\begin{split}
    &\begin{cases}
        \frac{1}{C_f} = \frac{1}{C_{th} \Delta T} \\
        -\frac{P_{loss}(t)}{C_f} = -\frac{\mathcal{Q}(t, T(t), \mathbf{x}_{env}(t))}{C_{th} \Delta T}
    \end{cases} \\
    \implies &
    \begin{cases}
        C_f = C_{th} \cdot (T_{max} - T_{min}) \\
        P_{loss}(t) = \mathcal{Q}(t, T(t), \mathbf{x}_{env}(t))
    \end{cases}
\end{split}
\label{eq:parameter_identification}
\end{equation}

Equation \eqref{eq:parameter_identification} not only concludes the proof but also fundamentally reveals the physical significance of the virtual parameters. 

Since $C_{th}$ is an intrinsic physical property of the building mass and $\Delta T$ is a fixed scalar defined by user comfort, $C_f$ is strictly a time-invariant scalar. It represents the total thermal energy storage capacity within the comfort range, providing a stable capacity metric for power grid dispatch. Conversely, since the heat leakage function $\mathcal{Q}$ depends dynamically on time-varying environmental states $\mathbf{x}_{env}(t)$ (such as weather fluctuations) and the system state $T(t)$, $P_{loss}(t)$ must be a time-variant parameter. It effectively acts as the instantaneous thermal load or the dynamic "leakage" rate of the virtual battery.

To further elucidate this theoretical mapping, consider the widely adopted first-order Equivalent Thermal Parameter (1R-1C) model as a specific instance. In the 1R-1C model, the thermal dynamics of the building envelope are characterized by a single lumped thermal resistance $R_{th}$ and capacitance $C_{th}$. Under this assumption, the generalized heat leakage function $\mathcal{Q}$ is linearly approximated based on Newton's law of cooling:
\begin{equation}
    \mathcal{Q}(t, T(t), \mathbf{x}_{env}(t)) = \frac{T_{out}(t) - T(t)}{R_{th}},
    \label{eq:1r1c_Q}
\end{equation}
where $T_{out}(t)$ is the outdoor ambient temperature. Substituting Eq. \eqref{eq:1r1c_Q} into the derived identification conditions in Eq. \eqref{eq:parameter_identification}, the virtual battery parameters for a 1R-1C AC system are explicitly instantiated as:
\begin{equation}
\begin{split}
    C_f &= C_{th} \cdot (T_{max} - T_{min}), \\
    P_{loss}(t) &= \frac{T_{out}(t) - T(t)}{R_{th}}.
\end{split}
\label{eq:1r1c_parameters}
\end{equation}
This specific example clearly demonstrates that while $C_f$ remains a static parameter determined by the building's inherent heat capacity and user settings, $P_{loss}(t)$ explicitly manifests as a time-varying variable driven proportionally by the indoor-outdoor temperature gradient.

Thus, the existence and properties of the parameters are proved. The virtual battery model is mathematically isomorphic to the generalized thermodynamic model under the derived parameter definitions. This rigorously justifies our motivation to employ data-driven methods to identify these physical parameters directly from operational data.

\section{VB-NET Design}
\label{network}
To bridge the gap between data-driven flexibility and physical interpretability, we propose VB-NET, a physics-constrained gray-box deep learning framework. Unlike conventional black-box models that directly map inputs to the target state (SOC), VB-NET is designed to solve the inverse problem of thermodynamic identification. The network does not predict the SOC directly; rather, it identifies the unknown physical parameters---the static equivalent capacity ($C_f$) and the time-varying power loss ($P_{loss}(t)$)---from historical operation data. These identified parameters are then fed into a differentiable physics layer to evolve the system state.

\begin{figure*}[htbp] 
\centering
\includegraphics[width=0.8\textwidth]{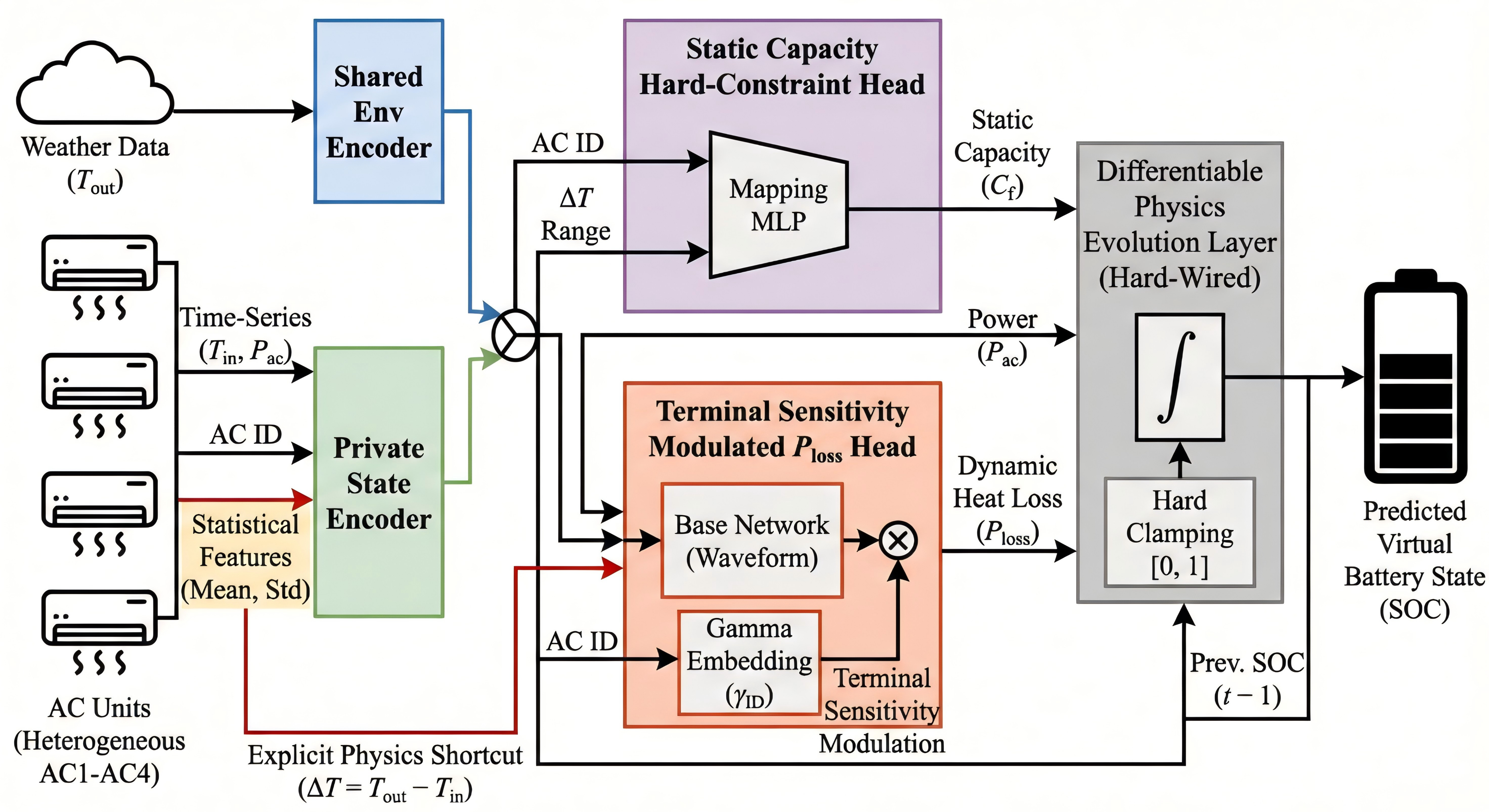}
\caption{The structure of VB-NET.}
\label{fig:architecture}
\end{figure*}

The overall architecture, illustrated in Fig. \ref{fig:architecture}, is composed of three distinct stages: (1) Disentangled Feature Encoding, (2) Physics-Informed Parameter Identification, and (3) Differentiable Physics Evolution. This design ensures that the model output strictly adheres to the energy conservation law while maintaining high adaptability to heterogeneous building characteristics.

\subsection{Disentangled Feature Encoding Mechanism}
Building energy dynamics are driven by two distinct forces: shared meteorological conditions and private building thermal characteristics. To capture this, we design a dual-stream encoding mechanism.

Acknowledging that buildings within the same district are subject to identical meteorological forcing, we employ a shared 1D Convolutional Neural Network (CNN) to extract environmental features. The encoder $E_{shared}$ maps the inputs to a latent representation vector $\mathbf{h}_{env}$:
\begin{equation}
    \mathbf{h}_{env} = E_{shared}(\mathbf{x}_{env}; \theta_{shared}),
\end{equation}
where $\mathbf{h}_{env}$ denotes the extracted high-dimensional environmental feature vector, $\mathbf{x}_{env} \in \mathbb{R}^{T}$ represents the time-series input of outdoor temperature, and $\theta_{shared}$ refers to the trainable network weights that are shared across all AC units to capture common weather patterns.

To characterize the heterogeneity of individual AC units, we introduce a private branch $E_{private}$. A critical innovation here is the explicit injection of statistical features as an inductive bias. The private encoder fuses dynamic states with static descriptors:
\begin{equation}
    \mathbf{h}_{state} = E_{private}([\mathbf{x}_{tin}, \mathbf{x}_{power}, \mu_{Tin}, \sigma_{Tin}, \mathbf{e}_{ID}]; \theta_{private}).
\end{equation}
Here, $\mathbf{h}_{state}$ is the encoded private feature vector for a specific AC  unit. The inputs include the indoor temperature sequence $\mathbf{x}_{tin}$, the power consumption sequence $\mathbf{x}_{power}$, and two statistical indicators: $\mu_{Tin}$ (mean indoor temperature) and $\sigma_{Tin}$ (standard deviation of indoor temperature), which serve as data-driven proxies for thermal resistance. $\mathbf{e}_{ID}$ represents the learnable identity embedding vector for the specific unit, and $\theta_{private}$ denotes the parameters of the private Multi-Layer Perceptron (MLP).

The outputs of both encoders are then concatenated to form the fused feature vector $\mathbf{h}_{fused} = [\mathbf{h}_{env}, \mathbf{h}_{state}]$, containing a complete description of both external drivers and internal thermal fingerprints.

\subsection{Physics-Informed Parameter Identification}
A core contribution of VB-NET is the separation of parameter identification into static and dynamic pathways, effectively decoupling the time-invariant capacity from time-varying losses.

According to our derivation, the virtual capacity $C_f$ depends solely on the building mass and the user-defined temperature deadband, making it a time-invariant scalar. We utilize a dedicated static head to predict this value:
\begin{equation}
    \hat{C}_f = \sigma \left( \mathcal{F}_{cap}(\mathbf{e}_{ID}, \Delta T_{range}) \right) \cdot (C_{max} - C_{min}) + C_{min}.
\end{equation}
Here, $\hat{C}_f$ is the estimated virtual capacity. $\sigma(\cdot)$ denotes the Sigmoid activation function ensuring the output is normalized between 0 and 1. $\mathcal{F}_{cap}$ represents the mapping function of the capacity head network. The inputs are the unit's ID embedding $\mathbf{e}_{ID}$ and the temperature deadband $\Delta T_{range} = T_{max} - T_{min}$. Finally, $C_{max}$ and $C_{min}$ are the upper and lower bounds of the capacity search space, used to scale the normalized output back to the physical domain.

The energy dissipation $P_{loss}(t)$ is highly dynamic and non-linear. To model this, we propose a Terminal Sensitivity Modulation mechanism, inspired by Feature-wise Linear Modulation (FiLM) \cite{perez2018film}. 

First, we establish a physics shortcut by explicitly concatenating the physical driving force $\Delta T_{phy}(t) = T_{out}(t) - T_{in}(t)$ with the latent features. This is grounded in the First Law of Thermodynamics, where heat flux is proportional to the temperature difference. A Base Network then learns a generalized heat loss waveform $\mathcal{P}_{base}$ driven by environmental conditions.

To adapt this generalized waveform to specific buildings (e.g., varying wall insulation), we introduce a learnable sensitivity scalar $\gamma_k$ for each AC unit $k$. The final predicted power loss is modulated as:
\begin{equation}
    \hat{P}_{loss}(t) = \mathcal{P}_{base}(\mathbf{h}_{fused}, \Delta T_{phy}) \cdot (1 + \gamma_k),
\end{equation}
where $\hat{P}_{loss}(t)$ is the estimated instantaneous power loss at time $t$. $\mathcal{P}_{base}$ is the Base Network that learns a generalized heat loss waveform. Its inputs are the fused features $\mathbf{h}_{fused}$ and the physical temperature difference $\Delta T_{phy}(t)$. The term $\gamma_k$ is a learnable scalar specific to the $k$-th AC unit, representing its unique sensitivity. The term $(1 + \gamma_k)$ acts as a scaling factor to modulate the base waveform for individual adaptation.

\subsection{Differentiable Physics Evolution Layer}

The final layer of VB-NET is a non-trainable, explicit implementation of the virtual battery dynamics. It acts as a differentiable solver:
\begin{equation}
    \hat{S}_{t+1} = \text{Clamp}\left( \hat{S}_t + \frac{\Delta t}{\hat{C}_f} \left[ \eta P_{ac}(t) - \hat{P}_{loss}(t) \right], 0, 1 \right),
\end{equation}
where $\hat{S}_{t+1}$ and $\hat{S}_t$ represent the predicted SOC at the next and current time steps, respectively. $\text{Clamp}(\cdot, 0, 1)$ ensures the SOC remains physically bounded between 0\% and 100\%.  $\eta$ is the energy efficiency ratio of the AC unit. $P_{ac}(t)$ is the actual electrical power input, and $\hat{C}_f$ and $\hat{P}_{loss}(t)$ are the physics parameters identified by the previous layers.

To ensure both numerical accuracy and dynamic consistency, the network is trained using a composite loss function:
\begin{equation}
    \mathcal{L} = \frac{1}{N} \sum_{t=1}^{N} \left( ||\hat{S}_t - S_{true, t}||^2 + \lambda ||\Delta \hat{S}_t - \Delta S_{true, t}||^2 \right),
\end{equation}
where $\mathcal{L}$ represents the total loss value. $N$ denotes the number of samples in a batch. $S_{true, t}$ is the ground-truth SOC derived from temperature measurements. The term $\Delta \hat{S}_t = \hat{S}_t - \hat{S}_{t-1}$ is the predicted rate of change, and $\Delta S_{true, t}$ is the actual rate of change. $\lambda$ is a hyperparameter balancing the weight between the state value loss and the derivative loss, forcing the model to capture correct physical dynamics.

\section{Case Study}
\label{case_study}
\subsection{Experiment Setup}
\label{subsec:experiment_setup}

To evaluate the proposed VB-NET, we simulate heterogeneous AC resources. According to the literature \cite{pan2017feasible,lu2012evaluation}, building thermal capacitance $C$ generally ranges from $3599$ to $4.483 \times 10^9$ J/$^\circ$C, and thermal resistance $R$ ranges from $6.7 \times 10^{-3}$ to $1.208 \times 10^2$ $^\circ$C/kW. The specific parameter settings for our simulated units are detailed in Table \ref{tab:ac_parameters}. In this study, $C$ and the energy efficiency ratio $\eta$ are uniformly fixed at $1.8 \times 10^7$ J/$^\circ$C and $0.97$ for all AC units. We design two experimental scenarios: \textbf{Case A} validates the overall feasibility of VB-NET using four AC units (AC1--AC4), while \textbf{Case B} tests the model's cold-start performance using sets of four (AC1--AC4) and eight (AC1--AC8) AC units.

\begin{table}[!htbp]
\caption{Settings of AC resources.}
\label{tab:ac_parameters}
\centering
\begin{tabular}{lccc}
\toprule
AC & $R$ ($^\circ$C/kW) & $\bar{P}$ (kW) & \textbf{$[\underline{T}, \bar{T}]\left({ }^{\circ} \mathrm{C}\right)$} \\
\midrule
AC1 & 3.0 & 12 & [21, 24] \\
AC2 & 3.5 & 12 & [22, 24] \\
AC3 & 5.0 & 13 & [21, 24] \\
AC4 & 6.0 & 10 & [20, 23] \\
AC5 & 5.0 & 12 & [21, 24] \\
AC6 & 5.5 & 11 & [22, 25] \\
AC7 & 6.5 & 10 & [20, 22] \\
AC8 & 6.0 & 12 & [20, 23] \\
\bottomrule
\end{tabular}
\end{table}

The hourly power consumption of each AC unit is generated using a price-responsive demand function \cite{gao2025deep}. The simulation is driven by real-world summer data (July 1st to September 30th, 2020--2023) from Shenzhen, China, utilizing real-time outdoor temperatures \cite{Shenzhentemperature} and local Time-of-Use electricity prices \cite{Shenzhenelectricity}. The generated dataset is chronologically divided into 80\% for training and 20\% for testing.

Regarding the specific configuration of VB-NET, the historical look-back window (sequence length) is set to 24 hours, and the latent hidden dimension is defined as 64. In the feature encoding stage, the Shared Environmental Encoder utilizes two 1D convolutional layers with 16 and 32 filters respectively (kernel size 3, padding 1), interleaved with Max-Pooling and ReLU activations. The Private State Encoder embeds the AC identity into an 8-dimensional vector and employs a Multi-Layer Perceptron (MLP) mapping the concatenated 35-dimensional input to the 64-dimensional latent space. For the parameter identification heads, the Static Capacity Head is formulated as an MLP with a 32-neuron hidden layer and a Sigmoid output scaled by the physical capacity bounds. The Terminal Modulated Power Loss Head utilizes a three-layer MLP (hidden sizes of 64 and 32) to generate the base loss waveform, which is subsequently scaled by a 1-dimensional learnable sensitivity embedding initialized at 0.5.

To quantitatively evaluate the predictive performance of the models, the Root Mean Square Error (RMSE) and the Coefficient of Determination ($R^2$) are adopted as the primary evaluation metrics. They are defined as follows:
\begin{equation}
    RMSE = \sqrt{\frac{1}{n}\sum_{i=1}^{n}\left(S_{true, i} - \hat{S}_i\right)^2},
\end{equation}
\begin{equation}
    R^2 = 1 - \frac{\sum_{i=1}^{n}\left(S_{true, i} - \hat{S}_i\right)^2}{\sum_{i=1}^{n}\left(S_{true, i} - \bar{S}_{true}\right)^2},
\end{equation}
where $n$ represents the total number of samples in the testing dataset. $S_{true, i}$ and $\hat{S}_i$ denote the ground-truth SOC and the predicted SOC at the $i$-th time step, respectively. $\bar{S}_{true}$ is the mean value of the actual SOC across all tested time steps. A lower RMSE and an $R^2$ closer to 1 indicate superior model accuracy.

All experiments were implemented in Python 3.8.8 with PyTorch 2.2.2 and CUDA 11.8, accelerated by an Intel\textsuperscript{\textregistered} Core\texttrademark{} i9-13980HX processor and an NVIDIA GeForce RTX 4070 Laptop GPU.

\subsection{Case A: Feasibility of VB-NET}
\label{subsec:case_a}

\begin{figure*}[htbp]
\centering
\includegraphics[width=1.0\linewidth]{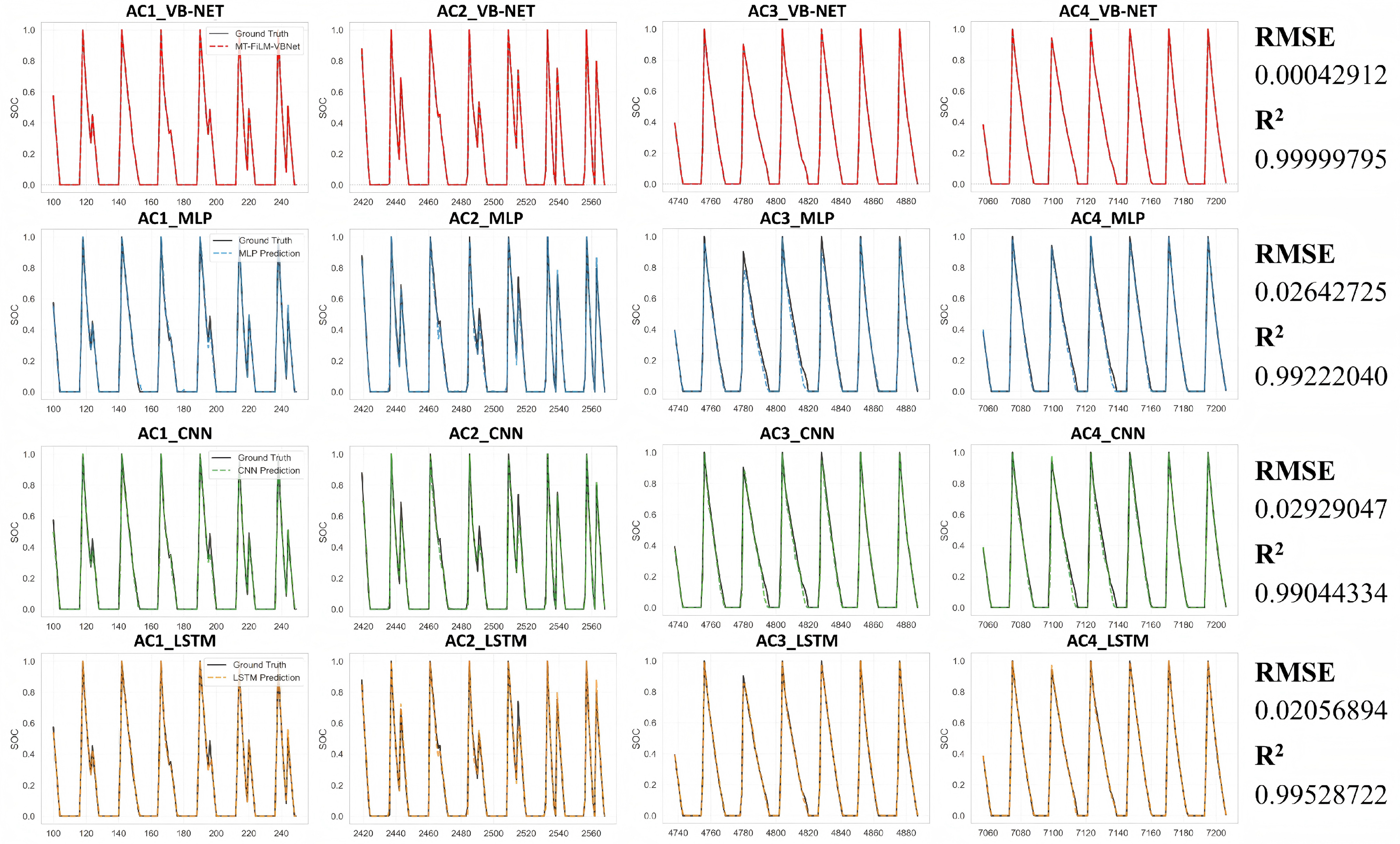} 
\caption{Comparison of SOC tracking performance between VB-NET and conventional deep learning models (MLP, CNN, LSTM).}
\label{fig:soc_tracking}
\end{figure*}

Fig. \ref{fig:soc_tracking} illustrates the SOC tracking performance of the proposed VB-NET compared to prevalent deep learning baselines, including MLP, CNN, and LSTM. While the baseline models struggle to accurately capture the sudden inflection points of the SOC curves due to their reliance on purely data-driven feature mapping without physical constraints, VB-NET demonstrates superior tracking capability. Evaluated by Root Mean Square Error (RMSE) and R-squared ($R^2$) metrics, VB-NET outperforms all baselines, with its predicted trajectory nearly overlapping the ground-truth SOC curve. This confirms the high numerical accuracy of VB-NET in abstracting AC resources into virtual batteries.

\begin{figure}[!htbp]
\centering
\includegraphics[width=0.7\linewidth]{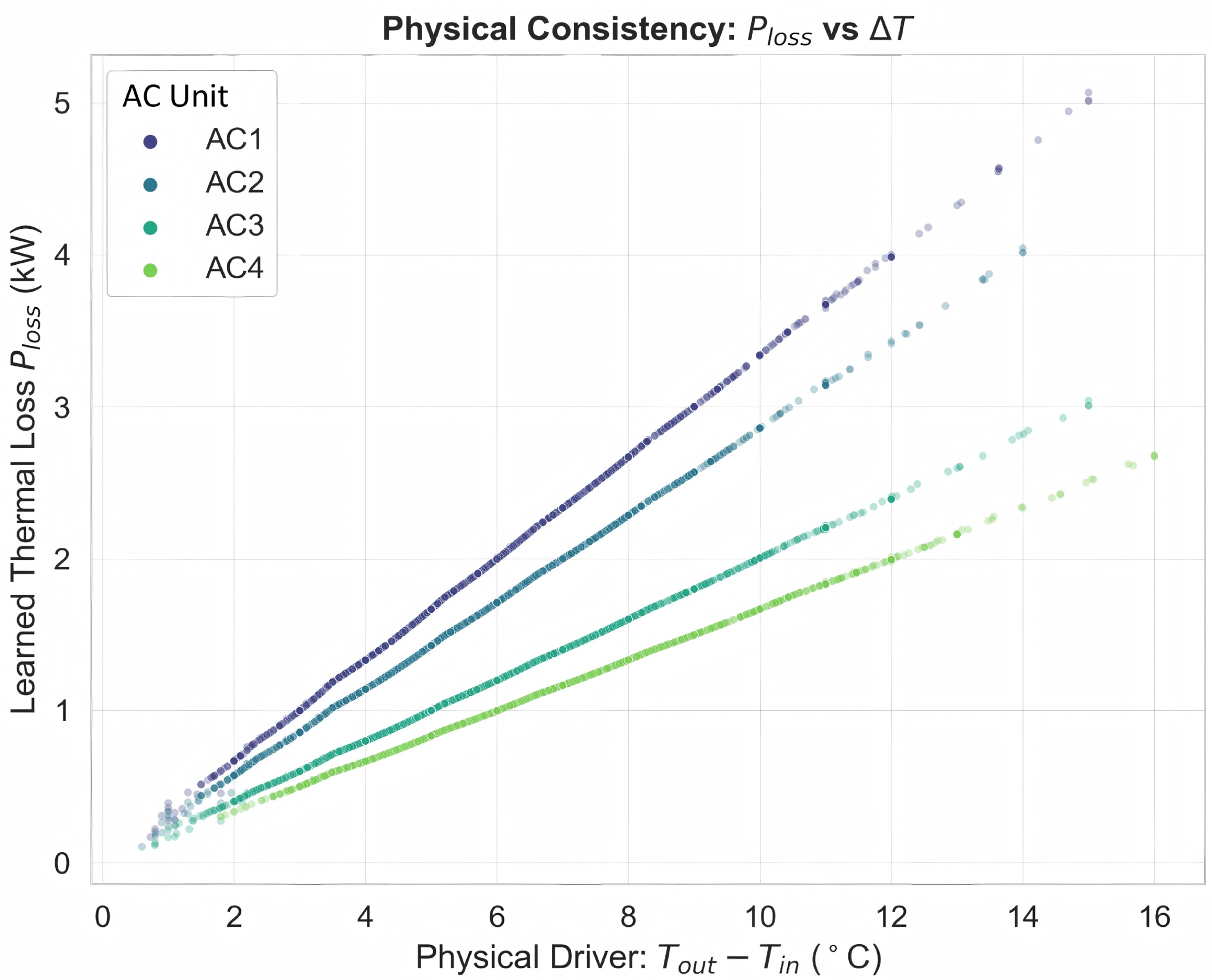} 
\caption{The linear relationship between the identified power loss ($P_{loss}$) and the indoor-outdoor temperature difference across different AC units.}
\label{fig:ploss_vs_temp}
\end{figure}

\begin{figure}[!htbp]
\centering
\includegraphics[width=0.7\linewidth]{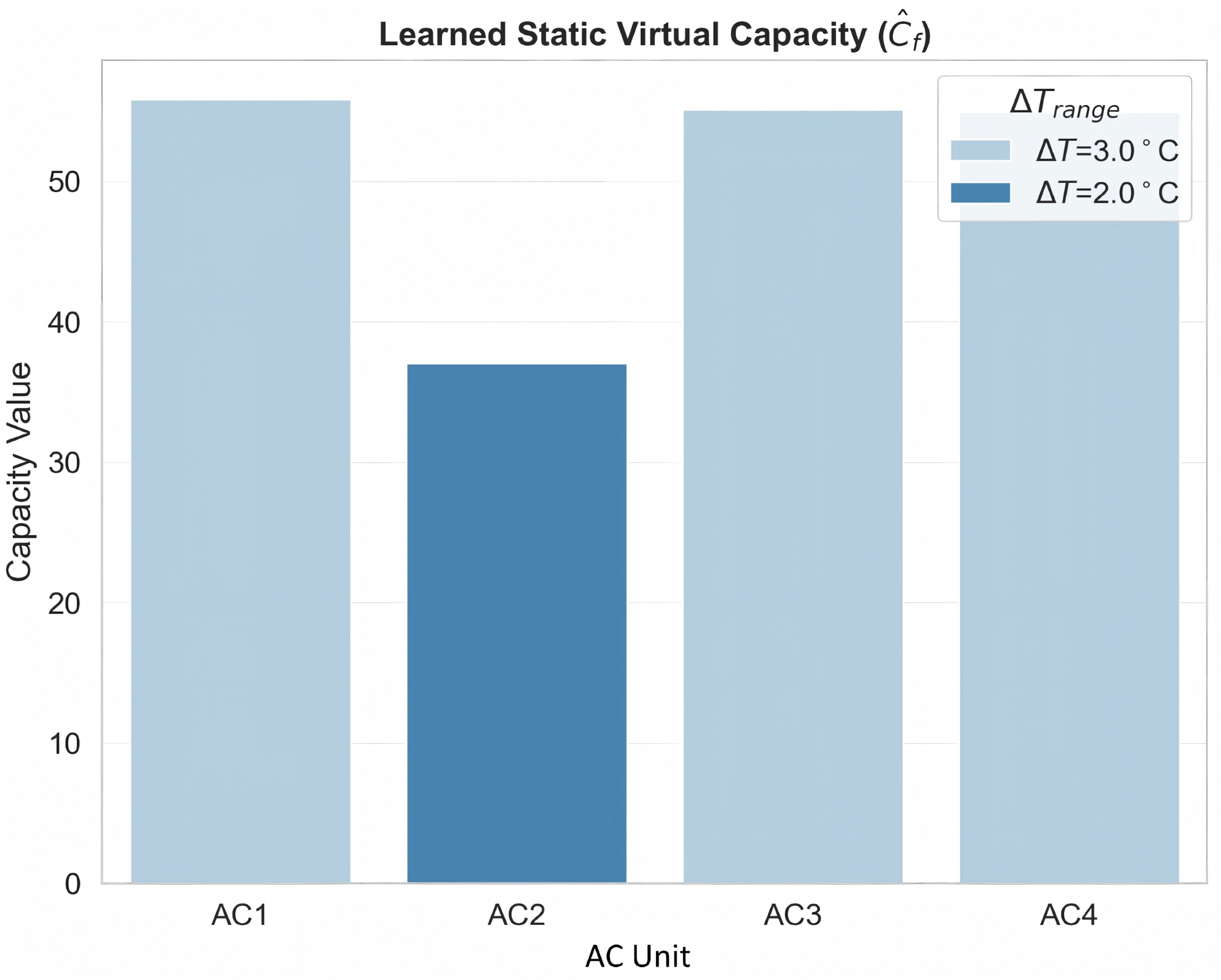} 
\caption{Comparison of the identified virtual capacity ($C_f$) under different temperature deadband settings.}
\label{fig:cf_comparison}
\end{figure}

Beyond numerical accuracy, VB-NET successfully recovers the underlying thermodynamic laws. According to Eq. \eqref{eq:1r1c_parameters} for a 1R-1C system, the power loss $P_{loss}$ is directly proportional to the indoor-outdoor temperature difference and inversely proportional to the thermal resistance $R$, the virtual capacity $C_f$ is proportional to the temperature deadband $\Delta T_{range}$. To validate this physical consistency, Fig. \ref{fig:ploss_vs_temp} and Fig. \ref{fig:cf_comparison} present the network's estimated $P_{loss}$ and $C_f$ under varying environmental conditions and temperature settings. As shown in Fig. \ref{fig:ploss_vs_temp}, the scatter points of $P_{loss}$ for different AC units form distinct linear trajectories, with slopes strictly inversely proportional to their respective $R$ values listed in Table \ref{tab:ac_parameters}. Furthermore, Fig. \ref{fig:cf_comparison} reveals that AC1, AC3, and AC4, which share a 3$^\circ$C temperature deadband, yield nearly identical $C_f$ estimations. In contrast, AC2, constrained by a narrower 2$^\circ$C deadband, exhibits a proportionally smaller $C_f$. These results demonstrate that VB-NET, despite its neural network foundation, transparently captures and enforces physical principles during parameter identification.

\begin{figure}[!htbp]
\centering
\includegraphics[width=0.7\linewidth]{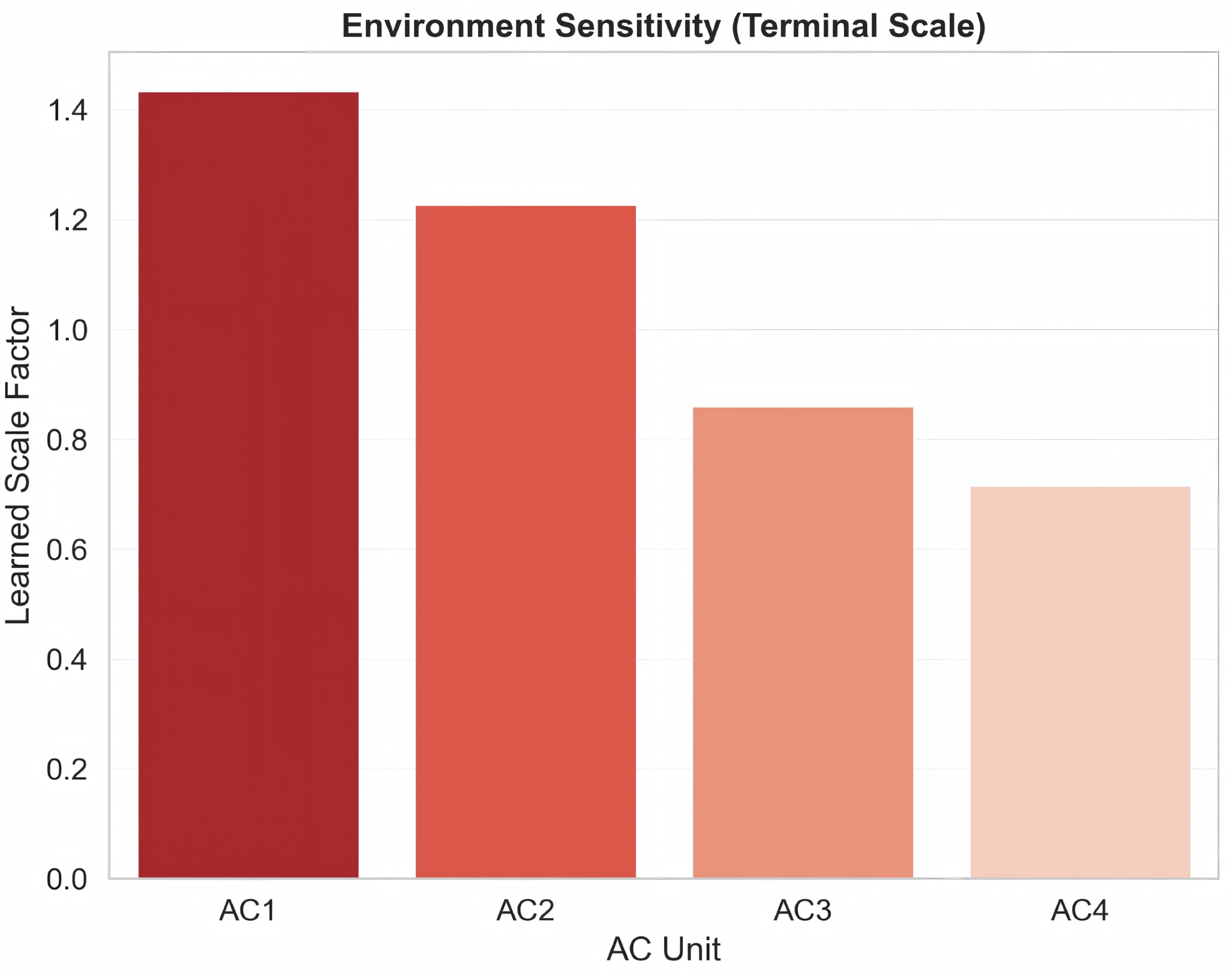}
\caption{The learned scale factors ($\gamma$) of the Terminal Sensitivity Modulation mechanism across different AC units.}
\label{fig:scale_factor}
\end{figure}

Furthermore, to validate the efficacy of the Terminal Sensitivity Modulation mechanism, we extracted the learned scale factors ($\gamma$) assigned to the different AC units, as depicted in Fig. \ref{fig:scale_factor}. Physically, this mechanism modulates the baseline power loss curve based on the building's susceptibility to environmental variations; a building more vulnerable to ambient changes should inherently exhibit a larger scale factor. From a thermodynamic perspective, this environmental sensitivity is inversely related to the building's thermal resistance ($R$). A lower $R$ value signifies higher thermal transmittance and, consequently, greater sensitivity to the outdoor environment. As detailed in Table \ref{tab:ac_parameters}, the thermal resistances of the four tested units follow the specific order of $R_1 < R_2 < R_3 < R_4$. Correspondingly, the learned scale factors shown in Fig. \ref{fig:scale_factor} perfectly align with this physical ordering, yielding a strict inverse correlation: $\gamma_1 > \gamma_2 > \gamma_3 > \gamma_4$. This exact alignment between the data-driven outputs and the underlying thermodynamic principles comprehensively demonstrates the physical consistency of the modulation module.

\begin{figure}[!htbp]
\centering
\includegraphics[width=0.8\linewidth]{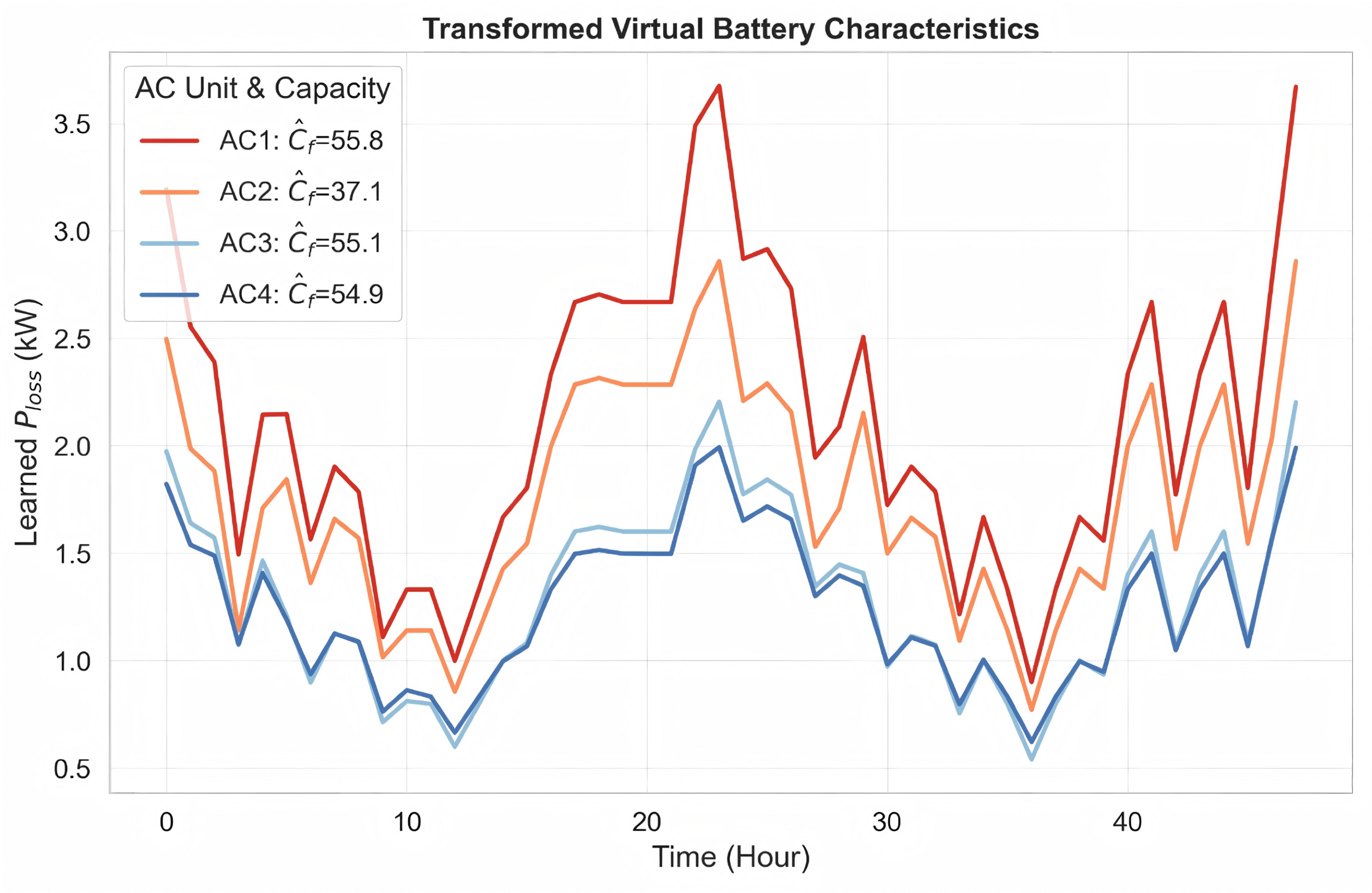} 
\caption{The identified time-varying virtual battery parameters ($P_{loss}$ and $C_f$) outputted by VB-NET for four individual AC units.}
\label{fig:time_varying_params}
\end{figure}

Finally, Fig. \ref{fig:time_varying_params} illustrates the time-varying virtual battery parameters generated by the model. The $P_{loss}$ trajectories of the four AC units exhibit highly consistent temporal trends, as they are identically driven by the same outdoor meteorological conditions and subsequently scaled by the Terminal Sensitivity Modulation mechanism. However, the specific magnitudes of the equivalent capacities and the fine-grained details of the loss curves vary across units, accurately reflecting their individual building thermal fingerprints. In summary, these findings comprehensively validate the accuracy, physical interpretability, and practical effectiveness of VB-NET in transforming heterogeneous AC systems into dispatchable virtual batteries.

\subsection{Case B: Cold-Start Experiment}
\label{subsec:case_b}

In this section, we conduct a cold-start experiment to evaluate the practical deployment capability of VB-NET. We simulate a realistic scenario where a newly integrated AC building in a given district possesses very limited historical operational data. Specifically, we investigate the performance difference between our Multi-Task Learning (MTL) architecture and a Single-Task Learning (STL) under varying availability ratios ($\alpha$) of the new unit's data. 

The STL baseline employs the identical VB-NET architecture but is trained from scratch exclusively on the available $\alpha\%$ data of the new AC, remaining completely isolated from other units. Conversely, the proposed MTL approach is evaluated under two joint-training configurations: a 4-AC setup (where AC1--AC3 serve as mature units utilizing 100\% data, and AC4 represents the new unit with $\alpha\%$ data) and an 8-AC setup (AC1--AC7 as mature units, AC8 as the new unit). The MTL models are jointly trained using 100\% of the mature units' data alongside the scarce $\alpha\%$ data of the new AC. This setup is designed to verify whether the Shared Environmental Encoder and the Terminal Sensitivity Modulation (FiLM) mechanism can successfully transfer generalized thermodynamic behaviors to the unseen unit. The comparative results are summarized in Table \ref{tab:cold_start}.

\begin{table}[!htbp]
\centering
\caption{Performance comparison (RMSE) of STL and MTL architectures under different data availability ratios ($\alpha$) for the newly integrated AC unit.}
\label{tab:cold_start}
\begin{tabular}{cccccc}
\toprule
\textbf{$\alpha$ (\%)} & \textbf{STL} & \textbf{MTL(3+1AC)} & \textbf{MTL(7+1AC)} \\
\midrule
2    & 0.072808  & 0.000343  & 0.012679  \\
4    & 0.072522  & 0.000463  & 0.000571  \\
6    & 0.072371  & 0.000448  & 0.000291  \\
8    & 0.072291  & 0.000479  & 0.000287  \\
10   & 0.072291  & 0.000343  & 0.000287  \\
25   & 0.072294  & 0.000367  & 0.000282  \\
50   & 0.002327  & 0.000382  & 0.000288  \\
100  & 0.000665  & 0.000429  & 0.000288  \\
\bottomrule
\end{tabular}
\end{table}

As shown in Table \ref{tab:cold_start}, the STL approach exhibits significantly degraded predictive performance under data-scarce conditions. For $\alpha \le 50\%$, the RMSE of the STL model stagnates at approximately 0.72, and acceptable precision is only achieved when the entire dataset ($\alpha = 100\%$) becomes available. This confirms that without the cross-validation and rich data distribution inherently provided by multi-task learning, the shared encoder and FiLM mechanisms in an isolated setup fail to abstract the universal environmental drivers, causing the network to easily fall into local optima.

In stark contrast, the MTL framework successfully achieves robust cold-start capabilities. In the 4-AC configuration, the MTL model essentially converges and matches the baseline 100\% data performance with merely $\alpha = 2\%$ of the new unit's data. In the 8-AC configuration, an extreme data imbalance at $\alpha = 2\%$ causes the model to over-prioritize the overwhelming volume of mature AC data, resulting in sub-optimal tracking for the new unit. However, as the new data ratio marginally increases, the 8-AC MTL model rapidly achieves high precision at $\alpha = 4\%$ and fully converges at $\alpha = 6\%$.

These findings demonstrate that despite the scarcity of private operational data for a new AC, the shared meteorological conditions act as a reliable bridge for knowledge transfer. By synergizing the Shared Environmental Encoder and the FiLM mechanism, VB-NET effectively extracts universal physical patterns and swiftly adapts them to new physical entities via minimal fine-tuning of the sensitivity scalar. 

\section{Conclusion}
\label{sec:conclusion}

This paper presented VB-NET, a novel physics-constrained deep learning framework designed to seamlessly translate air conditioning (AC) thermal dynamics into a dispatchable Virtual Battery (VB) model. Bridging the gap between rigorous thermodynamic modeling and flexible data-driven algorithms, we mathematically proved the isomorphism between the AC and VB models. To overcome the black-box nature of conventional neural networks, VB-NET incorporates a differentiable physics layer, strictly enforcing physical laws during the identification of the virtual capacity and time-varying power loss. 

Experimental validations demonstrated the superior performance and practical viability of the proposed framework. Compared to pure data-driven baselines such as MLP, CNN, and LSTM, VB-NET achieved state-of-the-art accuracy in tracking the equivalent SOC while eliminating trajectory deviations. More importantly, the network successfully extracted physically meaningful parameters; the identified time-varying power loss and static capacity strictly adhered to underlying thermodynamic relationships, such as the proportional dependence of thermal loss on environmental temperature gradients. Furthermore, to address the critical industrial challenge of deploying models for new buildings with scarce historical data, VB-NET incorporated a disentangled feature encoding mechanism alongside Terminal Sensitivity Modulation. Multi-task learning experiments verified that this mechanism effectively transfers universal meteorological driving patterns to unseen AC units, enabling rapid convergence and robust cold-start capabilities with as little as 2\% to 6\% of operational data. Ultimately, VB-NET offers a scalable and data-efficient solution for aggregating heterogeneous AC resources.

\bibliographystyle{elsarticle-num}  

\bibliography{cas-refs}



\end{document}